\documentclass[10pt]{article}
\usepackage{a4}
\usepackage{color}
\usepackage[english]{babel}
\usepackage{epsfig}
\usepackage[T2A]{fontenc}
\usepackage{amsfonts} 

\newcommand\Tr{{~\rm Tr~}}

\begin{document}
\title{Monte Carlo simulations\\ on Graphics Processing Units}
\author{Vadim Demchik\thanks{\tt E-mail: vadimdi@yahoo.com},
~~~ Alexei Strelchenko\thanks{\tt E-mail: alexstrelch@yahoo.com}
~\\~ {\small \sl{Dnepropetrovsk National University,
Dnepropetrovsk, Ukraine}}} \maketitle

\begin{abstract}
Implementation of basic local Monte-Carlo algorithms on ATI
Graphics Processing Units ($GPU$) is investigated. The Ising model
and pure $SU(2)$ gluodynamics simulations are realized with the
Compute Abstraction Layer ($CAL$) of ATI Stream environment using
the Metropolis and the heat-bath algorithms, respectively. We
present an analysis of both CAL programming model and the
efficiency of the corresponding simulation algorithms on GPU. In
particular, the significant performance speed-up of these
algorithms in comparison with serial execution is observed.
\end{abstract}


\section{Introduction}
In recent years designers of graphics hardware start the new
concept for games developers -- to accelerate game physics
directly on GPUs and, thus, to relieve less efficient for such
purposes Central Processor Units ({\it CPUs}) of {\it any} game
physics computations. Originally the GPU was the interface unit
between the system and monitor and has been optimized only for the
system memory - video memory data transferring and for storage of
videopages. The next evolutionary step of GPU was the appearance
of graphics accelerators. They were designed for simple and
frequently used video operations (displacement of large blocks of
images from one location to another on the screen, filling images,
and so on) without much usage of CPU. In the early and mid-1990s,
CPU-assisted real-time 3D graphics were becoming increasingly
common in personal computer and game consoles, which in turn lead
to ever-growing customer's needs for hardware-accelerated 3D
graphics. Over the past five years the computer graphics
technology has evolved from the possibility of performing basic
mathematical operations on image pixels to General Purpose
Computing on GPU technology ({\it GPGPU}).

Nowadays GPUs transformed into powerful programmable parallel
processors providing, in particular, double-precision floating
point support, which is especially essential for developing
scientific applications. As an illustrative example, the peak
performance of ATI RV 770 graphics processors exceeds 1 TFlops for
single-precision operations and is about 250 GFlops for
double-precision operations that is far beyond the performance
level of up-to-date high-end multicore CPUs. Another important
issue concerning GPU programming is availability of special
programming toolkits aimed at development of non-graphics
applications and allowing a programmer to abstract away from
sophisticated graphics application interfaces ({\it APIs}).
Corresponding tools, ATI Stream SDK and nVidia CUDA (Compute
Unified Device Architecture), can be freely downloaded from
official sites of leading GPU-makers. No wonder that GPGPU
attracts a lot of interest from the scientific community working
in such areas as astrophysics
(\cite{Ford:2008em}-\cite{Hamada:2007sw}), lattice QFT and spin
systems (\cite{Egri:2006zm}-\cite{Barros:2008rd}), numerical
analysis of PDE's \cite{Klockner:2009}, molecular and fluid
dynamics etc. (see, for instance \cite{gpgpu:2008} as well as
\cite{Collange:2009}-\cite{Shaw:2009} for further references).
Although there exist several reports on the topic, most of them
are based on utilization of nVidia's CUDA toolkit while ATI
programming environment seems to be less popular in the
literature. With the present publication we start our numerical
simulations on GPU with ATI Stream SDK to fill up this gap.

One of the main aims of the present study is to implement existing
Monte-Carlo ({\it MC}) algorithms, widely employed for simulations
in lattice QFT and spin models, on GPU and to investigate their
effectiveness on such parallel systems
\cite{Sokal:1996}-\cite{Berg:2004}. Of course, one should always
keep in mind that efficiency of a stochastic process is also
strongly effected by autocorrelation times, which are different
for different simulation schemes. That is, even though some
particular algorithm is badly parallelized in principle, as
non-local algorithms like the Wolff or Swendsen-Wang cluster
algorithm, nevertheless it might be quite effective in comparison
with some local algorithm, such as Metropolis algorithm that
suffers from large autocorrelation time in the critical region
(see however Ref. \cite{Prokofev:2001}, where another local
algorithm, the so-called worm algorithm, was investigated, that
proved to be deprived of this drawback). However, we do not study
efficiency of non-local updating schemes on GPU here.

The paper organized as follows. The second section contains the
implementation details for two local algorithms: the  Metropolis
and the heat-bath algorithms. In subsection 2.1 the common program
model of GPU-oriented simulation application is described. The
features of used pseudo-random number generator is presented in
subsection 2.2. Next subsection includes the description and
performance results of $2d$ Ising model simulations for Metropolis
algorithm on GPU while subsection 2.4 contains some details and
performance results of $SU(2)$ gluodynamics simulations for
heat-bath algorithm. In section 3 we discuss the obtained
performance results and perspectives for prospective
investigations. A short introduction into ATI Stream Technology is
shown in the Appendix A. General information about ATI RV770
hardware is collected in the Appendix B. Some basic details of ATI
Stream SDK is outlined in the Appendix C. Finally, in the Appendix
D we provide information on the hardware and software
configuration used in our Monte-Carlo simulations.

\section{Implementation}
We considered $2d$ Ising model and $SU(2)$ gluodynamics as a
perfect tool for testing various simulation algorithms. In this
section we examine implementation of the Metropolis and the
heat-bath updating algorithms on GPU for these two particular
models, respectively.

ATI Stream SDK allows to realize the algorithms by two software
tools: Brook+ and CAL (for general information see Appendix C). We
started to explore the ATI Stream technology with the Brook+, but
we found out quickly that its usage does not allow to write GPGPU
applications in the most efficient way. Brook+ is a very
transparent and suitable language for stream programming, but it
does not afford to implement some subtle points, needed for
optimal programming on GPU. In particular, we employed compute
shader model that provides more flexibility for general purpose
GPU computations. The basic ingredient of the compute shader is a
{\it thread} (while pixel shaders operate with {\it pixels}), one
invocation of a GPU computing program (or {\it kernel})
corresponding to a single element in the domain of execution. All
threads initiated by the kernel are composed in thread groups and,
within such a group, can communicate with each other via special
hardware unit, Local Data Share, assigned to each SIMD core. For
further details we refer to \cite{ATIr700} and \cite{ATIIL}.) The
critical requirement for performance is another essential argument
in favor of the CAL use. The present research is done with the
CAL.

\subsection{Program model}
Both the Ising model and $SU(2)$ gluodynamics simulations are
based on the same program model. As it is pointed in Appendix C,
ATI Stream SDK does not support the execution of a kernel from
another kernel. So, each kernel must perform a complete operation.
All kernels are run by the host program.

\begin{figure}
\begin{center}
\resizebox{0.5\textwidth}{!}{\includegraphics{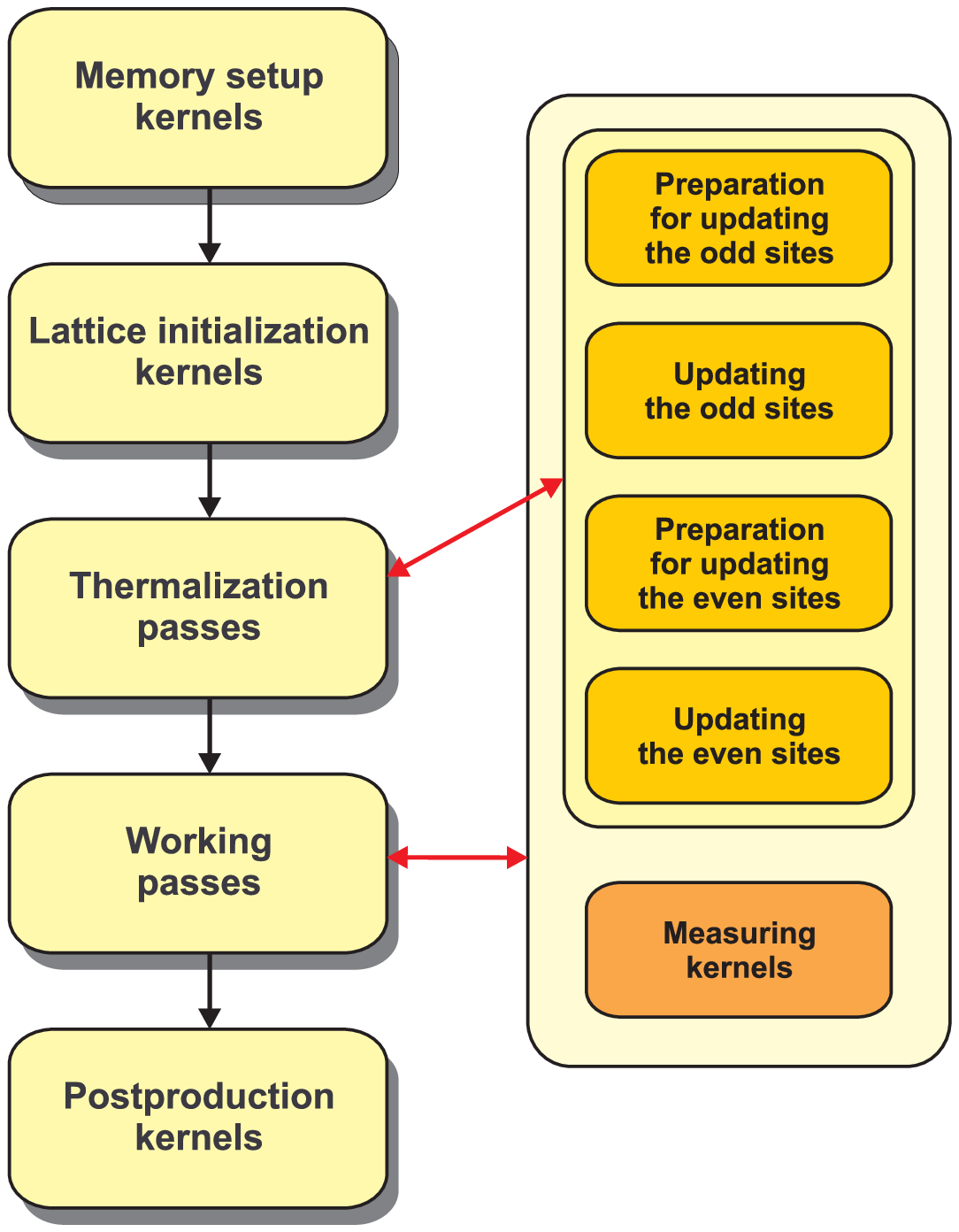}}
\label{fig:1} ~\\~\\~\noindent Figure 1:
\mbox{\parbox[t]{0.84\hsize}{\textsf{Structure scheme of
simulation program.}}}
\end{center}
\end{figure}

Another feature of the employed program model is that all
necessary data for simulations are stored in GPU memory. GPU
carries out intermediate actions and returns the results to host
program for final data handling and output. We avoid any data
transfer during run-time between the host program and kernels to
speed-up the execution process.

The structure scheme of simulation program is presented in Fig. 1.

First of all we have to allocate GPU memory. The most convenient
structure for this is the so-called {\it global buffer}, memory
space containing the arbitrary address locations to which uncashed
kernel outputs are written. ATI Stream SDK provides only one
global buffer per kernel, therefore, all kernels must utilize the
same global buffer to store all data. In addition, the memory
allocation may be accompanied by setup of initial values for some
internal variables in the global buffer. This is realized by
preparing necessary data arrays on host system memory with
subsequent mapping them into global buffer.

The next execution stage is lattice initialization. Two standard
initial states may be produced by the lattice initialization
kernels: ordered ({\it cold}) or disordered ({\it hot})
configuration. Lattice initialization may be obviously done by
host program before the memory setup kernel, however we carried
out full lattice simulation process entirely on GPU.

Thermalization and statistics sweeps are almost the same except
for the measuring kernels, which are not involved in the
thermalization process. The locality of the examined here
algorithms implies, in particular, that the lattice can be divided
into odd and even sites. To update odd sites, only information
from the even site is necessary and vice versa. Hence, the
complete MC sweep for all lattice requires the two groups of
kernels execution (to update the odd and even sites,
respectively). In working passes after each sweep the measuring
kernels can be run in order to measure microstate parameters
(microstate energy, correlations, etc).

Last stage is the execution of postproduction kernels
(such as statistical data handling). Afterwards, the export
of all data to the host memory must be done.

\subsection{Pseudo-random number generator}
In our simulations we employed the 24-bit Marseglia pseudo-random
number generator RANMAR\footnote{A rigorous analysis for a number
of parallel pseudo-random generators can be found in
\cite{Srinivasan:2003}.} \cite{Marsaglia}. Here we used the
implementation by James \cite{James:1988vf}. The algorithm is a
combination of a lagged Fibonacci generator and a simple
arithmetic sequence for the prime modulus $2^{24} - 3 = 16777213$.
The total period of RANMAR is about $2^{144} \approx 2.23 \times
10^{43}$.

The generator must be initialized by two given 5-digit integers
({\it initial seeds}), each set of which gives rise to an
independent sequence of sufficient length for an entire
calculation. The seed variables can have values between $0$ and
$31328$ for the first variable and $0$ and $30081$ for the second
variable, respectively. RANMAR can create, therefore, 900 million
independent subsequences for different initial seeds with each
subsequence having a length of approximately $10^{30}$ random
numbers. The 97 seed values and two offsets must be stored in GPU
global memory for every random number generator used in
simulations. So, it is reasonably to use a small number of random
number generators, which is smaller than number of lattice sites
for big lattices. As it was mentioned above, every GPU memory cell
has four 32-bit components, so it is convenient also to choose the
total number of generators to be multiplied by 4.

The generator consists of two parts: 1) kernel, which produces the
seed numbers on initial seed values; in fact this kernel is a
replica of RMARIN subroutine of James' version
\cite{James:1988vf}, and 2) subroutine, which directly produces
the random numbers. First part has been executed only for
initialization, and the second one has been running as internal
subroutine in all kernels that use random numbers (e.g. initial
lattice equilibration, MC updates). It is not necessary to store
generated random numbers in GPU memory in this case; such approach
essentially accelerates the run process.

In both Ising model and $SU(2)$ gluodynamics simulations we used
$4 \times 8192 = 32768$ generators, and their number may be chosen
in accordance with  the lattice size. The lowest 10 bits of
current thread number serve as identifier ({\it ID}) of random
number generator. It is possible to change easily the actual
number of generators by selecting any number of lowest bits in the
thread ID. This realization allows to start all generators as
independent parallel threads. In our implementation, RANMAR
generators produce integer random numbers, which are mapped into
the interval (of floating point values) $[0;1)$ afterwards.

\subsection{Metropolis algorithm for GPU $2d$ Ising model\\ simulations}
$2d$ Ising model, being the simplest and well-studied spin system,
is a handy toy-model for GPU programming experiments. We remind
the reader that subject to the character of the updating
procedure, all conventional dynamic Monte-Carlo algorithms fall
into two categories: namely, local and non-local algorithms. From
the parallel computing perspective, former category of MC
algorithms are usually the most convenient to deal with. Indeed,
typically the updating scheme in this case requires local
information only (e.g. information on close neighbors for a given
spin in the lattice etc.); as a consequence, all stream processors
work on their own distinct local data domain of the data with
little queuing or synchronization overheads. On the contrary, the
non-local algorithms are significantly harder to parallelize since
they demand much data communications for stream
processors\footnote{For instance, in cluster updating schemes the
main computational task is to take local information (e.g., bond
connections) and work out global information (i.e., cluster of
sites). Investigation of the parallel cluster algorithms can be
found in \cite{Apostolakis:1993}, \cite{Bae:1995}.}, the feature
is especially unwelcome for GPU programming. As it was already
noted, in the present study we are concerned about local MC update
procedures, and the simplest of them is the Metropolis algorithm.
Implementation of the Metropolis updates on parallel systems
presents no difficulties since one can always make use of standard
data decomposition of the lattice onto processors to achieve good
load balance. Here we use two-dimensional Ising model to
illustrate basic concepts of GPU parallel programming.

The partition function for the Ising model is defined as
\begin{equation}\label{one}
   Z(\beta) = \sum_{\{\sigma\}} \exp{(- \beta H)}, ~~~ H = - \sum_{<i,j>} \sigma_i \sigma_j
\end{equation}
with $\sigma_{i} = \pm 1$ being a spin variable at $i^{th}$ site
of the spin lattice. Here $< i, j >$ denote the summation over
nearest neighbor spins. It is quite natural to represent Ising
spins as 1-bit variables and to store them in registers of the
global buffer; in the case of GPU programming, this gives
additional advantage since it allows to reduce memory fetches as
well. The most essential part of the program is the spin lattice
partition. There are several ways to do this. The simplest one is
the checkerboard (or {\it ``red-black''}) partition, where each
stream processor has just one spin. All spin lattice is divided
into ``red'' and ``black'' squares, and all ``red'' (resp.,
``black'') spins are then processed in parallel. More efficient
way is to divide the system into small squares of some size and to
assign each stream processor to one such square. Besides, one has
also to take into account that the global buffer register is
four-component. Here we proposed the scheme that very much
resembles the well-known ``direct-dual'' decomposition of a spin
lattice; this is shown in Fig. 2. Black sites of the lattice
interact with ``dual'' red sites, and are composed into 4
component registers (in the figure it is indicated by gray painted
rectangles; each rectangle represents memory organization in the
global buffer). Similarly, ``red'' sites are collected in the
corresponding red painted rectangles  and thus stored in global
buffer registers. We found that such lattice organization allows
to reduce memory usage and better suites for GPU simulations.

\begin{figure}
\begin{center}
\resizebox{0.5\textwidth}{!}{\includegraphics{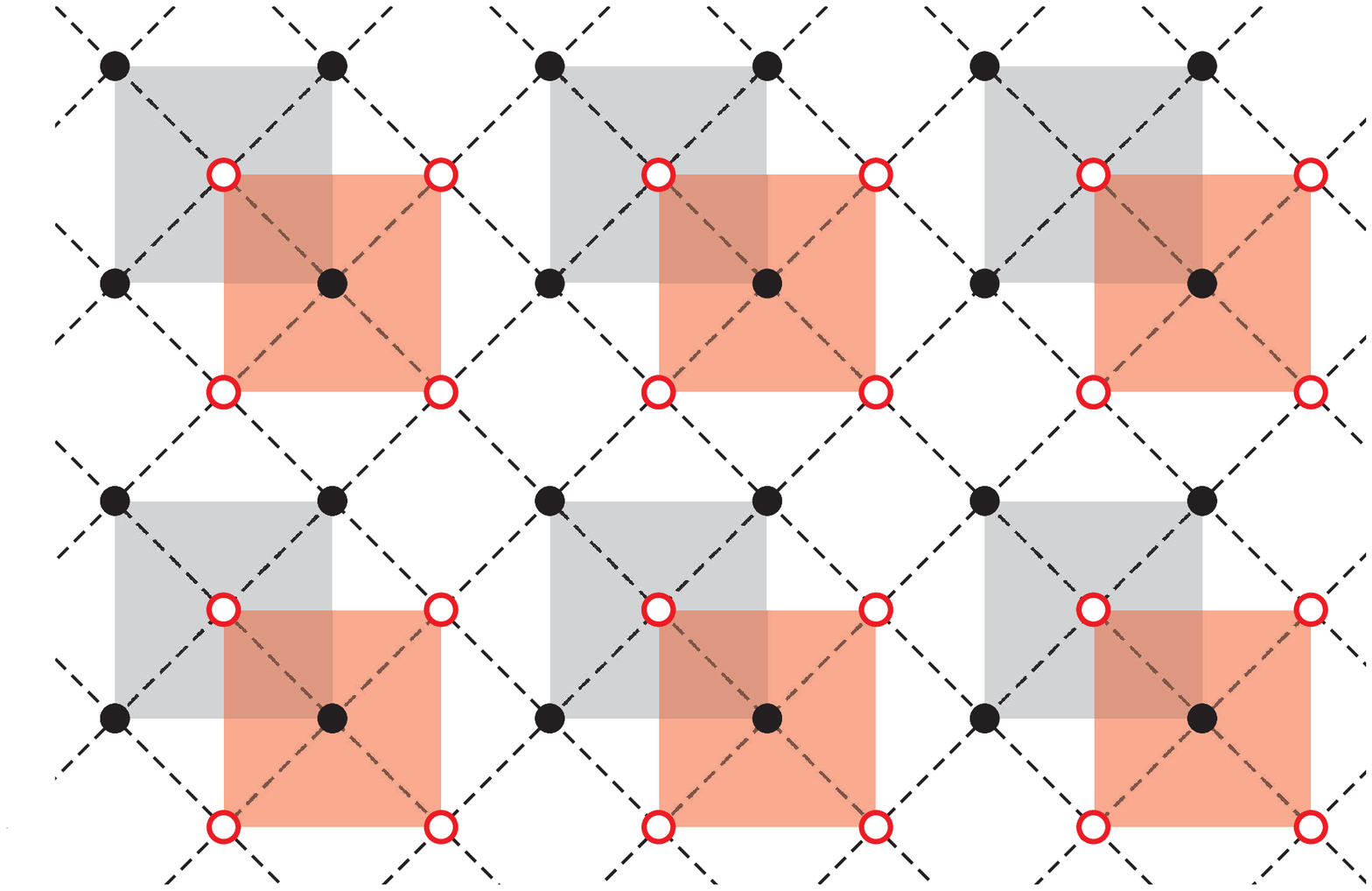}}
\label{fig:2} ~\\~\\~\noindent Figure 2:
\mbox{\parbox[t]{0.84\hsize}{\textsf{``Hybrid'' checkerboard
decomposition of the Ising lattice.}}}
\end{center}
\end{figure}
Simulation process on GPU was performed in the following stages:
\begin{enumerate}
  \item random number seed array preparation;
  \item loading of a chosen initial configuration for the spin
lattice in the GPU global buffer;
  \item equilibration run using the conventional Metropolis updating;
  \item statistics run;
\item finalization.
\end{enumerate}
These stages are actually common for both models (see below). In
the case of Ising model simulation, the source code contains eight
kernels. The first kernel is designed for generating seed number
arrays, which are stored in the GPU global buffer and accessible
for all remaining kernels. We also dedicate a single kernel for
the spin lattice initialization (as usual, it defines hot or cold
initial configuration). For equilibration run we use separate
kernels to perform ``even'' and ``odd'' sweeps, while statistic
runs are executed with the help of four kernels (two of them are
used for sweeps and measurements, and the other two perform
reduction operations and store obtained results for chosen lattice
observables in the global memory). In the present study we
consider lattice of medium size $1024 \times 512 = 524288$. In
particular,  we found that it takes about $630$ sec for $10^6$
lattice updates during equilibration phase and $730$ sec. for
$10^6$ measurements during statistic run phase.

\subsection{$SU(2)$ gluodynamics simulations}
In our $SU(2)$ gluodynamics simulations we used the hypercubic
lattice $L_t\times L_s^3$ ($L_t<L_s$) with hypertorus geometry;
$L_t$ and $L_s$ are the temporal and the spatial sizes of the
lattice, respectively. It should be noted that implemented algorithm
makes possible to work not only with spatial symmetric lattices
($L_1=L_2=L_3=L_s$), but also with totally asymmetric lattices
($L_1\not=L_2\not=L_3\not=L_t$).

The total number of lattice sites is $N_{sites}=L_1\times
L_2\times L_3\times L_t$. As it was mentioned, we divided whole
lattice into 2 parts - odd and even sites. The number of odd sites
is $N_{vect}=N_{sites}/2$, as well as the number of even ones.

The lattice may be initialized by cold or hot configuration. There
are 2 possibilities for hot initialization to fill lattice with
the pseudo-random $SU(2)$ matrices: constant series (by predefined
seeds) or random series (by system timer-dependent random seeds) of
pseudo-random numbers. In first case every startup process would
reproduce the same results for every execution (it is convenient
for debugging purposes).

Standard Wilson one-plaquette action of the $SU(2)$ lattice gauge
theory was used,
\begin{eqnarray}
S_W=\beta\sum_x\sum_{\mu<\nu}\left[1-\frac{1}{2}\Tr {\bf U}_{\mu\nu}(x)\right];\\
\label{Umunu} {\bf U}_{\mu\nu}(x)={\bf U}_\mu(x){\bf
U}_\nu(x+a\hat{\mu}){\bf U}^\dag_\mu(x+a\hat{\nu}){\bf
U}^\dag_\nu(x),
\end{eqnarray}
where $\beta=4/g^2$ is the lattice coupling constant, $g$ is the
bare coupling, ${\bf U}_\mu(x)$ is the link variable located on
the link leaving the lattice site $x$ in the $\mu$ direction,
${\bf U}_{\mu\nu}(x)$ is the ordered product of the link variables
(here and below we omitted the group indices).

The link variables ${\bf U}_\mu(x)$ are the $SU(2)$ matrices and
can be decomposed in terms of the unity, $I$, and Pauli
$\sigma_j$, matrices in the color space,
\begin{eqnarray}
{\bf U}_\mu(x)=IU^0_\mu(x)+i\sigma_jU^j_\mu(x).
\end{eqnarray}

Each global buffer cell comprises four 32-bit numbers (single
precision). Thus, it is natural to store all components of link
matrices ($U^0_\mu(x)$, $U^1_\mu(x)$, $U^2_\mu(x)$ and
$U^3_\mu(x)$) as one cell. For example, if general purpose
register is $R1={\bf U}_\mu(x)$, then the components of this register
are $R1.x=U^0_\mu(x)$, $R1.y=U^1_\mu(x)$, $R1.z=U^2_\mu(x)$ and
$R1.w=U^3_\mu(x)$.

To update the lattice, heat-bath algorithm with overrelaxation was
used \cite{Creutz:1987xi}, \cite{Ilgenfritz:1985dz}. At every MC
iteration we successively replaced each lattice link matrix with
\begin{eqnarray} \label{MCupd}
{\bf U}_\mu(x)\to\bar{{\bf U}}_\mu(x)=w {\bf W}_{\mu}(x),
\end{eqnarray}
where
\begin{eqnarray} \label{MCupdW}
{\bf W}_{\mu}(x)=\sum_{\nu ~(\mu<\nu)} &\Big[&{\bf U}_\nu(x){\bf
U}_\mu(x+a\hat{\nu}){\bf U}^\dag_\nu(x+a\hat{\mu})\\\nonumber
&&+{\bf U}^\dag_\nu(x-a\hat{\nu}){\bf U}_\mu(x-a\hat{\nu}){\bf
U}_\nu(x+a(\hat{\mu}-\hat{\nu}))\Big],
\end{eqnarray}
$w$ is a normalization constant such that
\begin{eqnarray}
\bar{{\bf U}}_\mu(x)\in SU(2).
\end{eqnarray}

In order to thermalize the system we passed 1000 MC iterations for
every run. For measuring we used 1000 MC configurations (separated
by 5 updates). MC procedure contains four kernels (see Fig. 1):
\begin{enumerate}
    \item kernel for measuring the value ${\bf W}_{\mu}(x)$ in eqn.(\ref{MCupdW}) for odd
    sites;
    \item kernel for updating the odd site's links;
    \item kernel for measuring the value ${\bf W}_{\mu}(x)$ in eqn.(\ref{MCupdW}) for even
    sites;
    \item kernel for updating the even site's links.
\end{enumerate}
First two kernels are executed sequently for each $\mu$ direction.
After their execution last two kernels are started for each $\mu$
direction. Updating kernels have 8 attempts per start to modify
the link matrices.

The run-time parameters, such as numbers of thermalization and
working MC sweeps, number of link updates per one MC sweep, etc.,
can be easily set up before execution.

In our program model each thread operates only one lattice site.
So, the stream for every kernel has $N_{vect}$ elements. It is
necessary to store $N_{vect}$ values of ${\bf W}_{\mu}(x)$ in global
buffer during the MC procedure. Odd and even sites used the
same region in global buffer for any $\mu$.

Working MC passes contain the measuring kernels, which calculate
microstate energy. All averaging measurements were performed with
the double precision. Every thread in measuring kernel produces
two 64-bit values of local averages (\ref{Umunu}) -- for spatial
and temporal plaquettes, correspondingly. These values are stored
in global buffer and summed up by so-called {\it reduction
kernel}, which decreases the dimensionality of a stream by folding
along one axis. Reduction kernel does not use the GPU power in the
best way, because at last execution stages less threads are run
than it could be. Nevertheless, such kernels must be used to avoid
the high memory consumption. The final double-precision values of
temporal and spatial averages are stored in global buffer and
transferred after all MC sweeps to the host for postproduction
analysis.

\begin{center}
{\small \vskip 0.5cm \noindent\begin{tabular}{|l|c|c|c|c|} \hline
~ & $4\times 8^3$ & $4\times 16^3$ & $4\times 24^3$ & $4\times
32^3$ \\\hline CPU, sec. & 30 & 274 & 1985 & 3796\\\hline GPU,
sec. & 5.2 & 17.9 & 52.8 & 120.6\\\hline GPU utilization & 50\% &
92\% & 97\% & 98\%\\\hline Speed-up, times & 5.8 & 15.3 & 37.6 &
31.5\\\hline
\end{tabular}
\vskip 0.1cm ~\noindent\begin{tabular}{|l|c|c|c|} \hline ~ &
$8\times 16^3$ & $8\times 24^3$ & $8\times 32^3$ \\\hline CPU,
sec. & 744 & 2060 & 7920\\\hline GPU, sec. & 32.7 & 102.3 &
397.4\\\hline GPU utilization & 95\% & 98\% & 99\%\\\hline
Speed-up, times & 22.8 & 20.1 & 19.9\\\hline
\end{tabular}
}
\end{center}
{\small ~\\\noindent\textsf{\bf{Table 1:}}
\mbox{\parbox[t]{0.84\hsize}{\textsf{Performance comparison of
serial (CPU) and parallel (GPU) MC simulations for pure $SU(2)$
gluodynamics on some lattices.}}}}

 ~\\~

The performance comparison of serial and parallel MC simulations
for different lattices is collected in Table 1. In the first row
the lattice geometry is shown. The next row contains the execution
time of Fortran-program, compiled with the trial version of Intel
Fortran compiler 11 (last available version) \cite{IVFC}. The
software and hardware configurations of simulation machines are
presented in Appendix D. The execution time of GPU-oriented
program is shown in the third row. The fourth row comprises
percentage of GPU utilization during the program execution. For
smallest lattice ($4\times 8^3$) this value is low because of low
density of kernel operations in the whole execution process. The total
speed-up of parallel (GPU) MC simulations in comparison with
serial (CPU) one is shown in the last row. It can be seen that for
small lattices ($4\times 8^3$, $4\times 16^3$) the gain in
productivity is not too high (5.8 and 15.3 times, respectively).
But for bigger lattices it varies from 19.9 to 37.6 times. We
believe that in full QCD simulations the total speed-up should be
much higher, because of big number of arithmetic operations per
every MC sweep.

\section{Discussion}
The main purpose of the present work is, on the one hand, to
demonstrate how MC simulations can be realized on GPU, and, on the
other hand, to estimate performance gain of implemented algorithms
with ATI Stream SDK. Here we investigated two basic local updating
schemes, the Metropolis and heat-bath algorithms, by example of
simplest physical systems ($2d$ Ising and $SU(2)$ gluodynamics,
respectively). We found significant performance increase of $10-50
\times$ in GPU numerical simulations in comparison with their
execution on contemporary CPUs. The corresponding source codes can
be sent via e-mails by request (see front page for e-mail
addresses).

Let us point out the key features of the implementation presented
here. First of all, it should be noted that we utilized a beta
version of ATI Stream SDK, in particular, some useful features
(such as LDS) were not realized to the full extent. For instance,
to avoid unstable behavior of the program, we were forced to
perform reduction via global buffer directly, though, it was more
natural to utilize special designed hardware unit (LDS) for this
purpose. Since this shortcoming might affect the applications
performance, it is very desirable, therefore, that it will be
fixed in future releases of ATI Stream SDK\footnote{During
preparation of this report ATI team released new version of their
programming toolkit, ATI Stream 1.4 beta. We did not test this new
version yet.}. The other obvious point is that our applications
can be run on ATI hardware only. However, recently approved
standard for Open Computing Language ({\it OpenCL}) will give an
opportunity to develop crossplatform applications, even for
heterogeneous computational systems (CPUs+GPUs). The support for
this new technology is already expected this year from almost all
leading hardware manufacturers \cite{Khronos}. As a particular
example, OpenCL was already included in MAC OS X Snow Leopard
\cite{MACX}. Finally, it should be emphasized that during the GPU
simulations the CPU is actually free, so the data postprocessing
may be combined with the next simulations. Moreover, the
programming model allows simultaneously use up to 8 GPU units per
host, so up to 8 points (for example, different initial lattice
parameters) may be obtained at one run.

One of the main directions for our future research is full QCD
simulations on GPU. As known, realization of fermionic sector on
the lattice is the most resource-intensive task, requiring
laborious matrix-algebra computations (see for recent review Ref.
\cite{Ishikawa:2008pf} and references therein). A task of such
kind is especially welcome for GPU implementation and some
discussion of this topic can be already found in the literature
\cite{Egri:2006zm}, \cite{Ibrahim:2008}. Indeed, one may expect an
essential growth of the performance in this case due to high
intensity of matrix arithmetic operations in such simulations with
relatively low memory import-export operations that are the
bottleneck for GPU kernels.

\appendix
\section{ATI Stream technology overview}
ATI Stream technology is a set of advanced hardware and software
technologies that enable ATI graphics processors, working in
concert with the CPU, to accelerate applications beyond just
graphics \cite{ATIstream}, \cite{ATIgsps}, \cite{ATIscto}.

\begin{figure}[h]
\begin{center}
\begin{minipage}[h]{0.45\linewidth}
\includegraphics[width=0.9\linewidth]{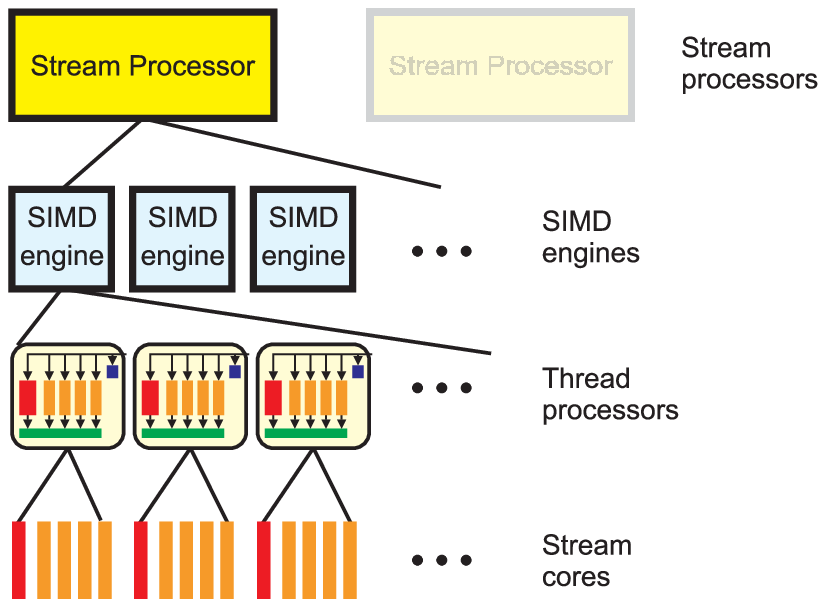}
\end{minipage}
\hfill
\begin{minipage}[h]{0.45\linewidth}
\includegraphics[width=0.9\linewidth]{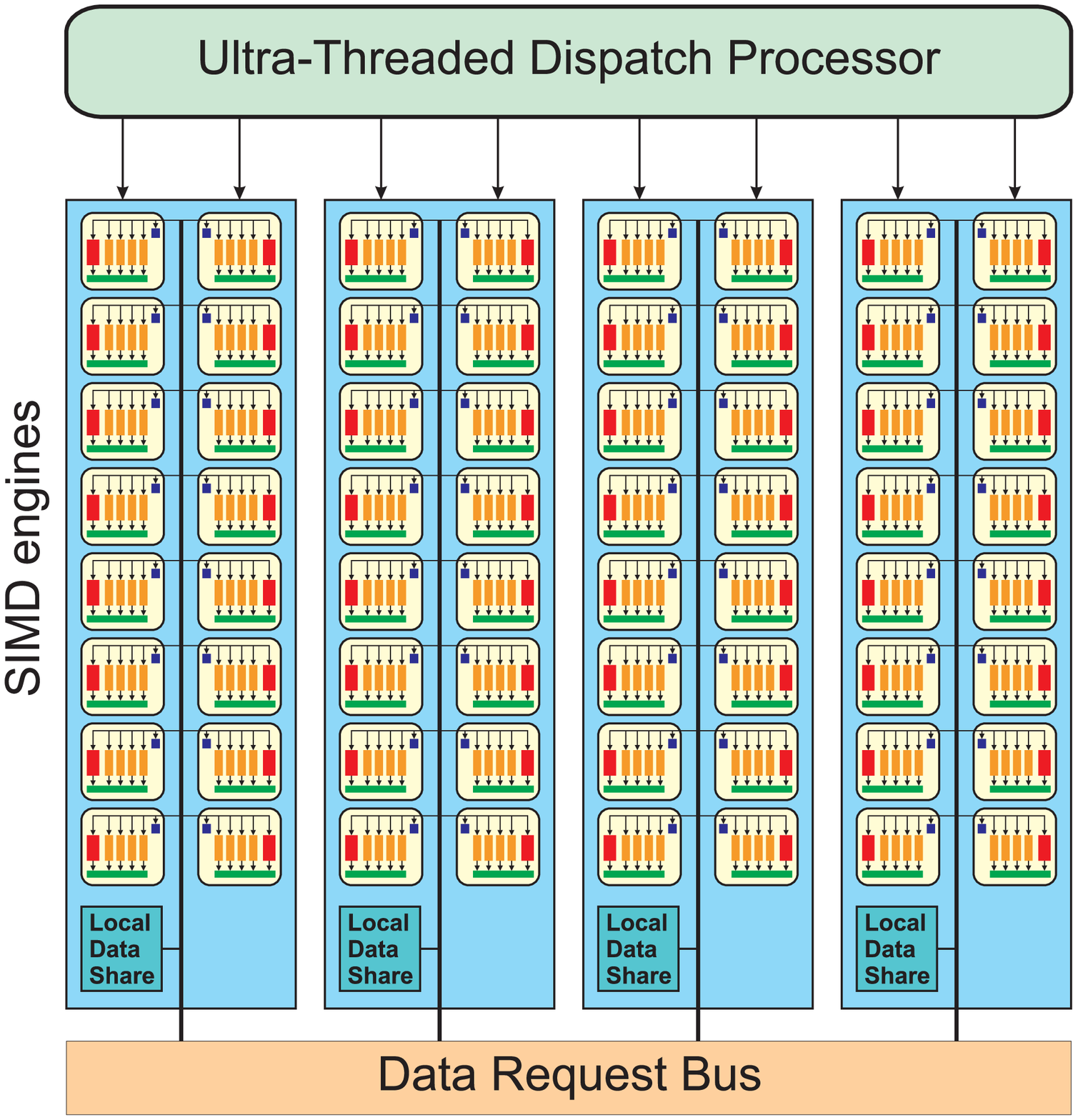}
\end{minipage}
~\\~\\~\noindent Figure 3:
\mbox{\parbox[t]{0.84\hsize}{\textsf{Generalized structure (left)
and simplified block diagram (right) of the stream processor.}}}
\end{center}
\end{figure}

A {\it stream} is a collection of data elements of the same type
that can be operated in parallel. A series of operations, which
are applied to each element in the stream, are called {\it stream
kernels} (or simply {\it kernels}). It is possible to run kernel
on some rectangular region of the stream (the so-called {\it
domain of execution}). {\it Stream processor} (or GPU) is the
hardware, which can operate the streams and execute kernels.

A modern graphics hardware can contain one or more stream
processors.  Different stream processors have different
characteristics, but follow a similar design pattern. In the Fig.
3 the generalized structure and simplified block diagram of stream
processor are presented. Each stream processor comprises of
numerous thread processors that are grouped into
single-instruction multiply-data ({\it SIMD}) engines (see
\cite{ATIgsps} and Fig. 3). Every SIMD engine contains the local
data share unit ({\it LDS}) aimed to fast transfer data between
the thread processors in a SIMD core. Thread processors contain
{\it stream cores}, which are the fundamental, programmable
computational units, responsible for performing integer, single
precision floating point, double precision floating point and
transcendental operations and the data types conversation. All
thread processors within a SIMD engine execute the same
instruction sequence; different SIMD engines can execute different
instructions. Each instance of a kernel running on SIMD engine's
thread processor is called a {\it thread}. The group of threads
that are executed together is called a {\it wavefront}. The number
of such threads in wavefront is called {\it wavefront size}. The
total execution time for the wavefront is determined by the thread
with the longest execution time.

\begin{figure}
\begin{center}
\resizebox{0.9\textwidth}{!}{\includegraphics{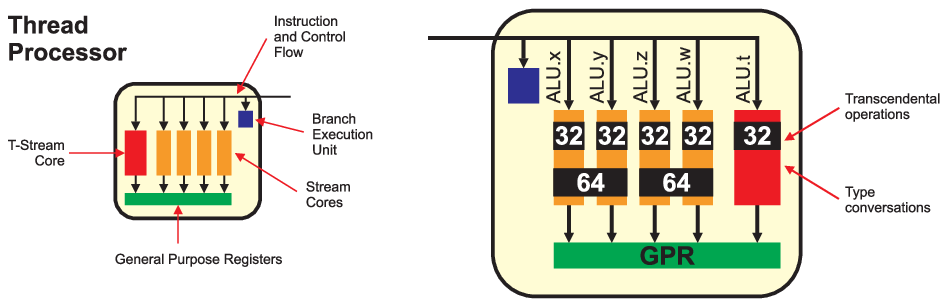}}
\label{fig:4} ~\\~\\~\noindent Figure 4:
\mbox{\parbox[t]{0.84\hsize}{\textsf{Simplified block diagram of
the VLIW thread processor.}}}
\end{center}
\end{figure}

A thread processor is arranged as a five-way VLIW processor (Fig.
4). Up to five scalar operations can be co-issued in a very long
instruction word ({\it VLIW}). To hide latencies due to memory
access and stream core operations, multiple threads are
interleaved. In a thread processor, up to four threads can issue
four VLIW instructions over four cycles. Stream cores can execute
single-precision floating point or integer operations. One of the
five stream cores ({\it T-stream core}) also can handle
transcendental operations. Double-precision floating point
operations are processed by connecting four of the stream cores
(excluding the transcendental core) to perform a single
double-precision operation. The thread processor also contains one
branch execution unit to handle branch instructions.

\section{General information about ATI RV770 videochip}
The most recent ATI Radeon architecture \textsf{RV770} was
released on June 25, 2008. There are several GPUs based on
\textsf{RV770} (\textsf{HD 4870}, \textsf{HD 4850}, etc.). The
\textsf{RV770} features a 256-bit memory controller and is the
first GPU to support GDDR5 memory (\textsf{HD 4870} and \textsf{HD
4870X2} cards are equipped with the GDDR5 memory), which possesses
a quadruple effective data transfer rate relative to its physical
clock rate (900 MHz*4=3600 MHz for \textsf{HD 4870} and \textsf{HD
4870X2}), instead of double as with GDDR3 memory. The bandwidth of
GDDR5 is up to 115 GB/s. All mentioned GPU cores are produced with
the 55 nm fabrication process.

ATI's \textsf{RV770} cards may be equipped with the 512, 1024 or
2048MB of memory. All mentioned engines are based on the system
bus PCI Express 2.0 x16, but also support PCI Express 1.1 and
1.0a.

The general information about some ATI's \textsf{RV770} video
cards is presented in Table 2. The last two columns are the
theoretical peak performance in single-precision floating point
operations (in Tflops) and thermal design power ({\it TDP}),
maximal value describing the thermal limits of cards (in Watts),
respectively.

{\small
\vskip 0.5cm
\begin{center}
\noindent\begin{tabular}{|l|c|c|c|c|c|c|}
\hline
Model & Core & Memory & Bandwidth & Bus width & Tflops & TDP \\
 & (MHz) & (MHz) & (GB/s) & (bit) & (peak) & (W) \\\hline
HD 4850 & 625 & 993 & 64 & 256 & 1.0 & 110 \\\hline
HD 4870 & 750 & 900* & 115.2 & 256 & 1.2 & 150 \\\hline
HD 4850X2 & 2x625 & 993 & 2x64 & 2x256 & 2.0 & 230 \\\hline
HD 4870X2 & 2x750 & 900* & 2x115.2 & 2x256 & 2.4 & 286 \\\hline
\end{tabular}
\end{center}
~\\\noindent\textsf{\bf{Table 2:}}
\mbox{\parbox[t]{0.84\hsize}{\textsf{General information about
some ATI's video cards \cite{wikiATI}.}}} ~\\~ }

Graphics cards \textsf{HD 4870X2} and \textsf{HD 4850X2} have two
stream processors each, the remaining \textsf{RV770} cards have
one stream processor only. Every stream processor of
\textsf{RV770} comprises 800 thread processors that are grouped
into 10 SIMD cores (see \cite{ATIgsps} and Fig. 3). The LDS size
for every \textsf{RV770} SIMD is 16KB. Thread processors of
\textsf{RV770} contain four stream cores and one T-stream core,
which are also named ALU.[x,y,z,w] and ALU.t, respectively (see
Fig. 4). Each pair of stream cores (i.e., ALU.x, ALU.y and ALU.z,
ALU.w) may be grouped into one 64-bit ALU for double-precision
operations. The \textsf{RV770} thread processor may co-issue up to
6 operations (five arithmetical/logical and one flow control
instruction). It also can perform 1.25 machine scalar operation
per clock for each of 64 data elements.

\section{ATI Stream SDK overview}
There are two main ATI Stream environments, which are composed
into one complete software development kit:
\begin{itemize}
    \item ATI Compute Abstraction Layer ({\it CAL}),
    \item Brook+.
\end{itemize}
ATI CAL is a device library that provides a forward-compatible
interface to ATI stream processors. ATI CAL lets software
developers interact with the stream processors cores at the
lowest-level for optimized performance, while maintaining forward
compatibility. ATI CAL is ideal for performance-sensitive
developers because it minimizes software overhead and provides
full-control over hardware-specific features that might not be
available with higher-level tools.

Brook is the high-level stream programming language for using
modern graphics hardware for non-graphical computations. Brook+ is
AMD's modified Brook open source data-parallel C++ compiler
\cite{ATIgsps}. Being built on top of ATI CAL, Brook+
source-to-source meta-compiler translates programs into
device-dependent kernels with the virtual instruction set
architecture ({\it ISA}) \cite{ATIr700}, the so-called ATI
Intermediate Language ({\it IL}). The generated C++ source
includes the CPU code and the stream processor device code, both
of which are later linked into the executable. Brook+ is a
higher-level language that is easier to use, but does not provide
all the functionality that ATI CAL does.

ATI IL is a pseudo-assembly language that can be used to describe
kernels for stream processors \cite{ATIIL}. It is designed for
efficient generalization of stream processor instructions, so that
programs can run on a variety of platforms without having to be
rewritten for each platform.

ATI CAL API comprises one or more stream processors connected to
one or more CPUs by a high-speed bus. The CPU runs the stream
processor device driver program and controls the stream processors
by sending commands using the ATI CAL API. Stream processor can
read from and write to its own local stream processor memory
and system memory using PCI Express bus. ATI CAL API exposes the
stream processors as SIMD array of computational processors.

The CAL runtime comprises system initialization and query, device,
context and memory management, program loading and execution.

In CAL all physical memory blocks allocated by the application for
the use in stream kernels are referred to as resources. These blocks
can be allocated as one-dimensional or as two-dimensional arrays
of data. There are several memory organizing structures, which may
be used with the CAL kernels: input buffers, output buffers,
constant buffers, scratch buffers, global buffers. The input
buffer can only be read from, not written to. The output buffer
can only be written to, not read from. Constant buffer is the
off-chip memory that contains constants. Scratch buffer is a
variable-sized space in off-chip memory that stores some of the
general purpose registers ({\it GPRs}). The global buffer lets
applications read from, and write to, arbitrary locations of
memory. Global buffers use a linear memory layout. All these
buffers support the following data types: signed or unsigned
32-bit integer, 32-bit floating point (single precision), 64-bit
floating point (double precision). A 128-bit address mapped memory
space consisting of four 32-bit components is called a register.
It is accepted to mark its components as .x, .y, .z and .w,
respectively. For double-precision operations the pair of 32-bit
components are grouped into 64-bit values (.xy and .zw).

ATI IL is a typeless language - the value in a register has no
intrinsic type, it is simply 32 or 64 bits of data \cite{ATIIL}.
Each mathematical instruction performs a typed computation --
signed or unsigned integer, floating point, double-precision
floating point -- on one or more untyped operands.

Unfortunately, there are some points, which make the programming
on GPU low-level SDK complicate. For instance, ATI CAL does not
support recursive execution of kernels as well as the execution of
a kernel from the other. Also, there are no runtime debugging
tools in ATI CAL. And finally, ATI Stream SDK still has beta
status, which may implies an occurrence of some compilation
errors.

Another strong point of ATI Stream SDK is the fact that ATI
drivers (ATI Catalyst) from version 8.12 contain all necessary
libraries for stream computing, so the compiled program could be
directly executed on any computer system equipped with the ATI
Radeon cards without the installation of additional software.

\section{Used hardware/software configuration}
\noindent Hardware configuration: {\small
\begin{itemize}
  \item {\it Ising model:} Intel Core 2 Duo CPU E6550 @ 2.33GHz, 4GB RAM memory;
  \item {\it SU(2):} Intel Core 2 Quad CPU Q6600 @ 2.40GHz, 4GB RAM memory.
\end{itemize}}

\noindent Graphics card specification: {\small
\begin{itemize}
  \item {\it Graphics card:} AMD/ATI Radeon HD 4850
  \item {\it GPU model:} RV770Pro
  \item {\it Number of thread processors:} 160 (combined in 10 SIMD cores)
  \item {\it Number of local memory fetch units:} 40 (4 for each SIMD core)
  \item {\it Engine clock:} 625 MHz
  \item {\it Memory clock:} 993MHz (GDDR3 type, effective memory clock -
1984MHz)
  \item {\it Memory bus:} 256 bit
  \item {\it Memory size:} 512MB
\end{itemize}}

\noindent Software configuration: {\small
\begin{itemize}
  \item {\it OS:} Windows XP SP3 (32-bit)
  \item {\it Used driver:} ATI Catalyst 9.1, \cite{ATIdrv}
  \item MS Visual Studio 2008 Express edition (C++ compiler), \cite{MSexp}
  \item ATI Stream SDK v. 1.3 beta, \cite{ATIsdk}
\end{itemize}
}

\section*{Acknowledgments}
One of the authors (V.D.) thanks E.-M. Ilgenfritz for kindly
provided Fortran program for $SU(2)$ lattice simulations.

\end{document}